\documentclass[journal=jacsat,manuscript=article]{achemso}

\usepackage[version=3]{mhchem} 
\usepackage{paralist}
\usepackage{booktabs,siunitx}
\usepackage{array}
\usepackage{dcolumn}
\newcommand{\mc}{\multicolumn}
\newcolumntype{H}{>{\setbox0=\hbox\bgroup}c<{\egroup}@{}}
\usepackage{amssymb}

\usepackage[flushleft]{threeparttable}
\usepackage[T1]{fontenc}
\usepackage{booktabs,braket,cleveref,mathptmx,xcolor,siunitx,dcolumn,graphicx,bm,graphics,caption,subcaption,multirow,array,paralist}
\usepackage[normalem]{ulem}%

\newcommand{\abinitio}{\emph{ab initio}}
\newcommand{\cm}{cm$^{-1}$}
\newcommand{\etal}{{\it et al}. }

\newcolumntype{d}{D{.}{.}{-1}}

\graphicspath{{Figs/}}

\author{Juan Camilo Zapata}
\author{Laura K. McKemmish}
\email{l.mckemmish@unsw.edu.au}
\affiliation[University of New South Wales]
{School of Chemistry, University of New South Wales, 2052, Sydney}

\title[An \textsf{achemso} demo]
  {VIBFREQ1295: A New Database for Vibrational Frequency Calculations}

\usepackage[utf8]{inputenc}
\DeclareUnicodeCharacter{2212}{-}
\begin{document}







\begin{abstract}
    High-throughput approaches for producing approximate vibrational spectral data for molecules of astrochemistry interest rely on harmonic frequency calculations using computational quantum chemistry. However, model chemistry recommendations (i.e, a level of theory and basis set pair) for these calculations are not yet available and, thus, thorough benchmarking against comprehensive benchmark databases is needed.
    
    Here, we present a new database for vibrational frequency calculations (VIBFREQ1295) storing 1,295 experimental fundamental frequencies and CCSD(T)(F12*)/cc-pVDZ-F12 \abinitio{} harmonic frequencies from 141 molecules. VIBFREQ1295's experimental data was complied through a comprehensive review of contemporary experimental data while the \abinitio{} data was computed here. The chemical space spanned by the molecules chosen is considered in depth and shown to have good representation of common organic functional groups and vibrational modes. 
    
    Scaling factors are routinely used to approximate the effect of anharmonicity and convert computed harmonic frequencies to predicted fundamental frequencies. With our experimental and high-level \abinitio{} data, we find that a single global uniform scaling factor of 0.9617(3) results in median differences of 15.9(5) \cm{}. Far superior performance with a median difference of 7.5(5) \cm{} can be obtained, however, by using separate scaling factors for three regions: frequencies less than 1000\,\cm{} (SF = 0.987(1)), between 1000 and 2000\,\cm{} (SF = 0.9727(6)) and above 2000\,\cm{} (SF = 0.9564(4)). This sets a lower bound for the performance that could be reliably obtained using scaling of harmonic frequencies calculations to predict experimental fundamental frequencies.
    
    VIBFREQ1295's most important purpose is to provide a robust database for benchmarking the performance of any vibrational frequency calculations.  VIBFREQ1295 data could also be used to train machine-learning models for the prediction of vibrational spectra, and as a reference and data starting point for more detailed spectroscopic modelling of particular molecules. The database can be found as part of the supplemental material for this paper, or in the Harvard DataVerse at \url{https://doi.org/10.7910/DVN/VLVNU7}.
\end{abstract}

\section{Introduction}
\label{sec:intro}

Definitive remote detection of molecular species in different astronomical settings, e.g (exo)planetary atmospheres, relies on the availability of high-resolution spectroscopic data \cite{90BeBeCr,99EnDrFe,01KrFe,08Ti,16McBrLo,12TeYu,22TeYu}, which is very time-intensive to produce. As a complementary approach, it is likely to be useful to obtain approximate data for the thousands of  molecular species of interest astrophysically, e.g. as discussed in \citet{16SeBaPe} Though this novel approach \cite{19SoPeSe,21ZaSyRo} cannot enable definitive molecular detections, there are nevertheless many anticipated applications of these big data including identification of molecules with strong absorption (that will be easier to detect), recognising potential ambiguities in molecular detections over some spectral windows, understanding the relative congestion of signals at different frequencies, and providing data for machine learning of vibrational frequencies. 

An obvious way to produce this approximate vibrational spectral data for thousands of molecules is through harmonic or anharmonic quantum chemistry vibrational frequency calculations, as piloted in \citet{21ZaSyRo} However, existing literature, as recently reviewed in Zapata Trujillo and McKemmish\cite{21ZaMc}, does not provide a definitive recommendation for a quantum chemistry methodology, even within the extremely popular harmonic approximation, nor explore the likely errors. These recommendations and error analysis are of interest far beyond the astrochemistry community (as evidenced by high citation numbers of, for example, Scott and Radom \cite{96ScRa}). We note that there has certainly been significant prior work on this topic (reviewed in Zapata Trujillo and McKemmish \cite{21ZaMc})  - it is the reliable definitive modern recommendations that are missing. Root-mean-squared errors (RMSE) between experimental and theoretical scaled harmonic frequencies have been computed extensively to compare model chemistry performance \cite{21ZaMc} but conclusions from this collation can only be preliminary as (1) the RMSE metrics cannot be fairly compared across different publications due to differences in the benchmark datasets used; and (2) there is often a lack of data for many important modern model chemistry choices.

The need for a new thorough and comprehensive benchmarking study of model chemistry performance for harmonic frequency calculations is clear \cite{21ZaMc}, but crucially the quality of this study relies on the quality of the experimental database against which theoretical results are compared. A key focus of this paper is to provide such an updated and collated experimental database, which we call VIBFREQ1295. The need for an update is justified initially by the fact that it has been almost thirty years since the most extensive and frequently used database was compiled in \citet{93PoScWo} 1064f/122mol (note this is widely known as the F1 set due to its use in Scott and Radom \cite{96ScRa}), then in hindsight by the significant number of updates to this original data that were found.

Given that using scaled harmonic frequencies is the most widespread approach to predict experimental fundamental frequencies in quantum chemistry, understanding the inherent errors of  the harmonic approximation at very high levels of model chemistry, e.g., CCSD(T)-like calculations, is pivotal for providing directions into less-demanding and routine calculations. For instance, such comparison can allow us to explore how best to quantify the error distribution and occurrence of large outliers in the calculations. Moreover, we can also examine different approaches for the implementation of scaling factors, highlighting errors and expected performance in each case. 

In the present paper, we provide a new benchmark database for vibrational frequencies, containing 1,295 experimental fundamental frequencies from 141 organic-like molecules (VIBFREQ1295). The frequencies collated here correspond to the most up-to-date experimental fundamental frequencies for the molecules considered, as recorded by an extensive literature review search. The uncertainties in the experimental measurements (when available), the symmetry of the vibration, as well as approximate descriptions to the vibrational modes, are also provided for all molecules in the database. To complement the experimental frequencies, we also calculated high-level CCSD(T)(F12*)/cc-pVDZ-F12 \abinitio{} harmonic frequencies for all molecules in the database. The reader can access the VIBFREQ1295 database in the supplemental material for this paper, as well as through the Harvard DataVerse platform \cite{Dataverse} (\url{https://doi.org/10.7910/DVN/VLVNU7}).

This paper is organised as follows. In Section \ref{sec:data_collation} we explain the criteria used to select the molecules and frequencies included in the database, and provide a comparison between the newly compiled experimental frequencies and the data from the previous vibrational frequency databases. A detailed analysis into the chemical space and functional groups included into VIBFREQ1295 is presented in Section \ref{sec:database_overview}, with particular interest in the distribution of different vibrational modes in the database. In Section \ref{sec:qc-calcs} we present the high-level (CCSD(T)(F12*)\cite{10HaTeKo}/cc-pVDZ-F12\cite{08PeAdWe,10HiPe}) \abinitio{} harmonic frequencies for the molecules together with our analysis in terms of the scaling factor calculation, its implication on the accuracy of the scaled harmonic frequencies, and the expected anharmonicity error. Finally, in Section \ref{sec:conclusions} we summarise our findings, outlining concluding remarks and future directions for this work. 

\section{Constructing the Database}
\label{sec:data_collation}

\begin{figure*}
    \centering
    \includegraphics[width=0.9\textwidth]{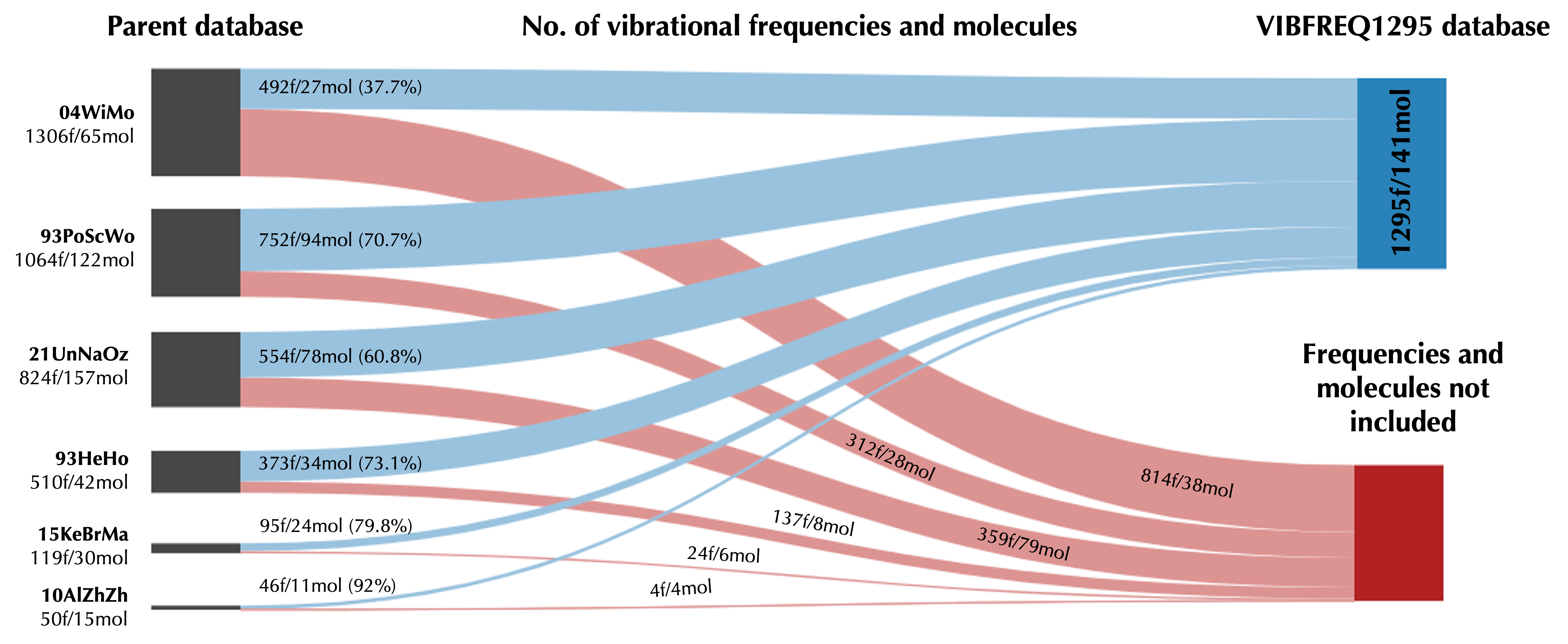}
    \caption{Vibrational frequency databases (left) used in the development of VIBFREQ1295 (upper-right blue panel). These databases correspond to the works of Witek and Morokuma \cite{04WiMo} (04WiMo - 1306f/65mol), \citet{93PoScWo} (93PoScWo -- 1064f/122mol a.k.a. the F1 set), \citet{21UnNaOz} (21UnNaOz -- 824f/157mol), Healy and Holder \cite{93HeHo} (93HeHo -- 510f/42mol), \citet{15KeBrMa} (15KeBrMa -- 119f/30mol), and \citet{10AlZhZh} (10AlZhZh -- 50f/14mol). The notation \textit{(x)f/(y)mol} indicates the total number of frequencies $x$ and molecules $y$ considered in each database. The same notation is used to indicate the number of frequencies and molecules from each database that were (upper-right blue panel) and were not (lower-right red panel) included into VIBFREQ1295.}
    \label{fig:datasets}
\end{figure*}

Herein, we present a new database for vibrational frequency calculations containing 1,295 experimental fundamental frequencies and high-quality (CCSD(T)(F12*)\cite{10HaTeKo}/cc-pVDZ-F12\cite{08PeAdWe,10HiPe}) \abinitio{} harmonic frequencies for 141 molecules. \Cref{tab:all_mol} presents an overview of the molecules considered in the database, showcasing the total number of atoms and experimental frequencies, frequency range covered, and references to the original experimental publications.

\begin{table}
\resizebox{15cm}{!}{
    \centering
    \caption{Summary of molecular species considered into VIBFREQ1295 listing the molecular formula, IUPAC name, total number of atoms and frequencies, the frequency range covered (in \cm{}), number of vibrational frequencies, and the references to the original experimental publications per molecule. Cyclic molecules are indicated with $c-$ before the molecular formula. Isomeric molecules (except for $cis$ and $trans$) are distinguished by an underscore and number after the molecular formula, e.g., \ce{C3H4} and \ce{C3H4}$\_$1. The electronic state for each molecular species is indicated as a superscript in the left-hand side of the molecular formula whenever the electronic state is not a singlet. Only $^{\textit{1}}$\ce{CH2} (singlet state) is explicitly presented in the table to distinguish it from $^{\textit{3}}$\ce{CH2} (triplet state).}
    \label{tab:all_mol}
    \begin{tabular}{llclclllclcl}
    \toprule
    \mc{1}{l}{Formula} & \mc{1}{l}{IUPAC Name} & \mc{1}{l}{\#At} & \mc{1}{l}{Freq. (\cm{})} & \mc{1}{l}{Vib. Freqs.} & \mc{1}{l}{Refs} & \mc{1}{l}{Formula} & \mc{1}{l}{IUPAC Name} & \mc{1}{l}{\#At} & \mc{1}{l}{Freq. (\cm{})} & \mc{1}{l}{Vib. Freqs.} & \mc{1}{l}{Refs} \\
    \midrule
        \ce{AlCl3}             & Aluminium chloride               & 4  & 616 - 151   & 6  & \cite{83ToSjKl}                                                         & \ce{ClNO}               & Nitrosyl chloride                         & 3  & 1799 - 331  & 3  & \cite{19Or, 88McKaGe} \\
        \ce{BH}                & $\lambda^{1}$-borane             & 2  & 2269 - 2269 & 1  & \cite{91FeBe}                                                           & \ce{ClNO2}              & Chloro nitrite                   & 4  & 1683 - 370  & 6  & \cite{94DuKiGu, 19FlAnKw, 20AnKwMa, 94OrMoGu, 98OrMoKl} \\
        \ce{BH3CO}             & Borane carbonyl                  & 6  & 2456 - 313  & 12 & \cite{57BeKeBe, 72LaPeCa, 76PeLaCa}                                     & \ce{$^{\textit{2}}$ClO} & Chlorosyl                                 & 2  & 853 - 853   & 1  & \cite{01DrMiCo} \\
        \ce{$^{\textit{2}}$BO} & Oxoboron                         & 2  & 1772 - 1772 & 1  & \cite{85MeDuBr}                                                         & \ce{$^{\textit{2}}$CN}  & Cyano radical                             & 2  & 2042 - 2042 & 1  & \cite{92PrBe} \\
        \ce{c-C2H4S3}          & 1,2,4-trithiolane                & 9  & 2976 - 82   & 21 & \cite{95DoGu}                                                           & \ce{CO}                 & Carbon monoxide                           & 2  & 2169 - 2169 & 1  & \cite{91Le}  \\
        \ce{c-C3H3NO}          & 1,3-oxazole                      & 8  & 3160 - 609  & 18 & \cite{07HeLaPa, 75MiPoSa}                                               & \ce{CO2}                & Carbon dioxide                            & 3  & 2349 - 667  & 4  & \cite{04MiBr, 82KaJoHo} \\
        \ce{c-C3H3NO}\_1       & 1,2-oxazole                      & 8  & 3160 - 591  & 18 & \cite{05Ro}                                                             & \ce{COCl2}              & Carbonyl dichloride                       & 4  & 1828 - 301  & 6  & \cite{15KwLaFl, 17FlKwLa, 16NdPeKw, 16FlTcLa} \\
        \ce{c-C3H6}            & Cyclopropane                     & 9  & 3101 - 737  & 21 & \cite{73BuJo, 09MaMaBl, 92PlMePi, 91MePlPi, 89PlTeLa}                   & \ce{COClF}              & Carbonyl chloride fluoride                & 4  & 1875 - 408  & 6  & \cite{01PeFlBu} \\
        \ce{c-C3H6S}           & Thietane                         & 10 & 2994 - 113  & 24 & \cite{88ShCaIb, 95DoGu}                                                 & \ce{COF2}               & Carbonyl difluoride                       & 4  & 1944 - 581  & 6  & \cite{97McDuMa, 13CoDrBr, 91CaPeFl} \\
        \ce{c-C4H4N2}          & Pyrazine                         & 10 & 3069 - 350  & 24 & \cite{94HeShBr}                                                         & \ce{CS}                 & Methanidylidynesulfanium                  & 2  & 1285 - 1285 & 1  & \cite{95RaBeDa}  \\
        \ce{c-C4H4O}           & Furan                            & 9  & 3169 - 599  & 21 & \cite{01MeAuDe, 94KlChSt, 93PaYaWi, 11ToCuBi, 99MeHeDe, 01MeLiHe}       & \ce{CS2}                & Carbon disulfide                          & 3  & 1535 - 395  & 4  & \cite{88WeScMa, 85BlBaCa, 80JoKa} \\
        \ce{c-C4H5N}           & 1H-pyrrole                       & 10 & 3530 - 474  & 24 & \cite{08HeWuPa2, 99MeVaHe, 95HeHe, 94KlChSt, 09ToWi, 01MeLiHe}          & \ce{CSCl2}              & Thiocarbonyl dichloride                   & 4  & 1138 - 297  & 6  & \cite{10McBi, 15McBi, 70FrBlBe} \\
        \ce{c-C4H8O2}          & 1,4-dioxane                      & 14 & 2970 - 274  & 36 & \cite{04WiMo}                                                           & \ce{CSF2}               & Difluoromethanethione                     & 4  & 1368 - 417  & 6  & \cite{72HoRuOv} \\
        \ce{c-C5H5N}           & Pyridine                         & 11 & 3094 - 373  & 27 & \cite{98Kl}                                                             & \ce{F2}                 & Molecular fluorine                        & 2  & 893 - 893   & 1  & \cite{94MaBeSa}  \\
        \ce{c-CH2N4}           & 1H-tetrazole                     & 7  & 3447 - 578  & 15 & \cite{00BiEnKe}                                                         & \ce{F2O}                & Fluoro hypofluorite                       & 3  & 928 - 460   & 3  & \cite{87BuSc, 86TaJoRu, 87Ta} \\
        \ce{C2Cl2}             & 1,2-dichloroethyne               & 4  & 2234 - 171  & 7  & \cite{92Mc, 70KlKlCh}                                                   & \ce{F2SO}               & Thionyl fluoride                          & 4  & 1334 - 376  & 6  & \cite{82OhNo, 03DaNeSm} \\
        \ce{C2H2}              & Acetylene                        & 4  & 3372 - 612  & 7  & \cite{93VaHuCa, 81HiKa, 72FaWe}                                         & \ce{FCN}                & Carbononitridic fluoride                  & 3  & 2318 - 451  & 4  & \cite{67CoIsLo, 97FaBrDu, 85JoLiTa} \\
        \ce{C2H2O}             & Ketene                           & 5  & 3165 - 439  & 9  & \cite{00NeLuQu, 87DuFeHa, 94EsDoCa, 87DuFeHa2, 72JoStWi}                & \ce{H2CO}               & Formaldehyde                              & 4  & 2843 - 1167 & 6  & \cite{06PeVaDa, 03PeKeFl} \\
        \ce{C2H2O2}            & Oxaldehyde                       & 6  & 2844 - 127  & 12 & \cite{11PrSaJo, 70CoOs, 02LaHeCe, 71CoOs}                               & \ce{H2O}                & Water                                     & 3  & 3755 - 1594 & 3  & \cite{56BeGaPl, 76CaFl} \\
        \ce{C2H3Cl}            & Chloroethene                     & 6  & 3129 - 395  & 12 & \cite{66EnAs, 00LoGiBi, 97GiStLo, 97StGiGh, 05DeMoPe, 92GiLoPe}         & \ce{H2S}                & Sulfane                                   & 3  & 2628 - 1182 & 3  & \cite{98BrCrCr, 96UlMaKo} \\
        \ce{C2H3N}             & Acetonitrile                     & 6  & 3040 - 362  & 12 & \cite{79PaPeAr}                                                         & \ce{HCl}                & Hydrochloric acid                         & 2  & 2990 - 2990 & 1  & \cite{99PaHiLe} \\
        \ce{C2H3N}\_1          & Isocyanomethane                  & 6  & 3013 - 260  & 12 & \cite{56Wi, 92HeBe, 95LePlGo, 96KhPaBr}                                 & \ce{HCN}                & Hydrogen cyanide                          & 3  & 3311 - 711  & 4  & \cite{00MaMeKl} \\
        \ce{C2H3OF}            & Acetyl fluoride                  & 7  & 3043 - 123  & 15 & \cite{72BeCo}                                                           & \ce{$^{\textit{2}}$HCO} & Formyl radical                            & 3  & 2434 - 1080 & 3  & \cite{88DaLaCu, 83BoDuLo, 78JoMcRi} \\
        \ce{C2H4O}             & Acetaldehyde                     & 7  & 3014 - 143  & 15 & \cite{71HoGu, 95AnTrPa, 99HeHeGe, 80Ho, 94KlHe}                         & \ce{HNCO}               & Isocyanic acid                            & 4  & 3538 - 577  & 6  & \cite{83StPoMc, 90YaWiJo, 97BrBeCr, 79StWiWi} \\
        \ce{C2H4O2}            & Methyl formate                   & 8  & 3045 - 130  & 18 & \cite{86ChHaMa}                                                         & \ce{HNO}                & Nitroxyl                                  & 3  & 2683 - 1500 & 3  & \cite{87PeVe, 77JoMc} \\
        \ce{C2H4O2}\_1         & Acetic acid                      & 8  & 3585 - 93   & 18 & \cite{65HaNo,05HaFeBl}                                                  & \ce{HNO3}               & Nitric acid          & 5  & 3550 - 458  & 9  & \cite{92TaLoLu, 96TaWaLo, 98Pe, 96LoLaWa, 13Pe, 04PeOrFl, 88GoBuHo} \\
        \ce{C2H5F}             & Fluoroethane                     & 8  & 3000 - 243  & 18 & \cite{20DiZiPo}                                                         & \ce{$^{\textit{2}}$HO2} & Hydroperoxy radical                       & 3  & 3436 - 1097 & 3  & \cite{98YaEnHi, 91NeZa, 92BuHaHo} \\
        \ce{C2H6}              & Ethane                           & 8  & 2985 - 289  & 18 & \cite{11LaLaVa, 08BoMoHo, 87HeFaKn, 11LaLaHo, 16MoOl}                   & \ce{HOCl}               & Hypochlorous acid                         & 3  & 3609 - 724  & 3  & \cite{86LaOl, 79WeSaLa} \\
        \ce{C2H6O}             & Methoxymethane                   & 9  & 2993 - 198  & 21 & \cite{16KuWeWa, 02CoCaCh, 64PeFoJo}                                     & \ce{HOF}                & Hypofluorous acid                         & 3  & 3577 - 889  & 3  & \cite{89BuPaRa, 88BuPaRa} \\
        \ce{C2H6S}             & Methylsulfanylmethane            & 9  & 2998 - 175  & 21 & \cite{94ElStGe, 18JaBeBe}                                               & \ce{N2F2}               & (E)-difluorodiazene                       & 4  & 1522 - 365  & 6  & \cite{66KiOv} \\
        \ce{C2HCl}             & Chloroethyne                     & 4  & 3339 - 324  & 7  & \cite{02HoRuWe, 02NeNiVa}                                               & \ce{N2H4}               & Hydrazine                                 & 6  & 3398 - 376  & 12 & \cite{52GiLi, 75DuGrMa, 60YaIcSh} \\
        \ce{C2HF}              & Fluoroethyne                     & 4  & 3356 - 366  & 7  & \cite{92HoNeMi}                                                         & \ce{N2O}                & Nitrous oxide                             & 3  & 2223 - 588  & 4  & \cite{14TiChCh, 07Ho, 18AlLaGa} \\
        \ce{C2N2}              & Oxalonitrile                     & 4  & 2330 - 233  & 7  & \cite{11Ma}                                                             & \ce{NCl2F}              & Diclhorofluoroamine                       & 4  & 825 - 274   & 6  & \cite{68HiAnHa} \\
        \ce{C3H3Cl}            & 3-chloroprop-1-yne               & 7  & 3335 - 186  & 15 & \cite{63EvNy}                                                           & \ce{NClF2}              & Chlorodifluoroamine                       & 4  & 930 - 378   & 6  & \cite{63Et, 66Co} \\
        \ce{C3H3F}             & 3-fluoroprop-1-yne               & 7  & 3338 - 211  & 15 & \cite{63EvNy}                                                           & \ce{NF3}                & Nitrogen trifluoride                      & 4  & 1032 - 493  & 6  & \cite{18BoUlBe, 15Na} \\
        \ce{C3H3N}             & Acrylonitrile                    & 7  & 3123 - 228  & 15 & \cite{15KiMaPi, 99KhNoPa}                                               & \ce{$^{\textit{3}}$NH}  & $\lambda^{1}$-azane                       & 2  & 3125 - 3125 & 1  & \cite{99RaBeHi} \\
        \ce{C3H4}              & Propa-1,2-diene                  & 7  & 3085 - 352  & 15 & \cite{02NiHeJo, 82PlMa, 14EsJoBe, 85MaPiDa, 86OhYaKu, 94He}             & \ce{NH3}                & Amonia                                    & 4  & 3443 - 968  & 6  & \cite{81UrSpPa, 99KlBrTa, 00CoKlTa} \\
        \ce{C3H4}\_1           & Prop-1-yne                       & 7  & 3335 - 330  & 15 & \cite{94KeLePa, 93GoCrPe, 04PrMuKl, 87HeTh, 14EsJoBe, 09PrMuUr, 85HeTh} & \ce{NHF2}               & Difluoroamine                             & 4  & 3193 - 500  & 6  & \cite{63CoMaSc} \\
        \ce{C3H4O}             & Prop-2-enal                      & 8  & 3103 - 157  & 18 & \cite{18McBiXu, 85HaNiTs, 15McBiXu, 07McToAp, 08McAp}                   & \ce{$^{\textit{2}}$NO2} & Nitrogen dioxide                          & 3  & 1616 - 749  & 3  & \cite{98PeFlGo} \\
        \ce{C3H6}              & Prop-1-ene                       & 9  & 3091 - 189  & 21 & \cite{94AiFrOr, 73SiLaPe, 06LaFlHe, 14EsAlFa, 18SuToDr}                 & \ce{NSCl}               & Azanylidyne(chloro)-$\lambda^{4}$-sulfane & 3  & 1325 - 271  & 3  & \cite{76MuMoCy, 06RoMc} \\
        \ce{C3H8}              & propane                          & 11 & 2976 - 217  & 27 & \cite{01FlLaHe, 65GaKi, 70GrPuLi, 10KwTcFl, 10FlTcLa, 19PeFlKw}         & \ce{NSF}                & Azanylidyne(fluoro)-$\lambda^{4}$-sulfane & 3  & 1372 - 366  & 3  & \cite{84MiEmDu} \\
        \ce{C3O2}              & Carbon suboxide                  & 5  & 2289 - 60   & 10 & \cite{78LoBr, 08WeMaPe, 79LoBr, 80FuMiGu, 94AuHoJe, 81HaMiKa, 98VaJoPo} & \ce{O3}                 & Ozone                                     & 3  & 1103 - 700  & 3  & \cite{02WaBiSc}  \\
        \ce{C4H2}              & Buta-1,3-diyne                   & 6  & 3333 - 219  & 13 & \cite{84GuCrRa, 92McBr}                                                 & \ce{OCS}                & Carbonyl sulfide                          & 3  & 2062 - 521  & 4  & \cite{06VaFa, 82Ka} \\
        \ce{C4H6}              & Buta-1,3-diene                   & 10 & 3100 - 162  & 24 & \cite{15KrCrBo, 04HaHaNe, 04CrDaHa, 08CrSa, 20MaPoGo}                   & \ce{ONF}                & Nitrosyl fluoride                         & 3  & 1844 - 519  & 3  & \cite{98EvMcDe, 84FoJo} \\
        \ce{C4N2}              & Dicyanoacetylene                 & 6  & 2296 - 107  & 13 & \cite{97WiSc, 94WiKeGu, 92WiKeNi, 92WiScLe}                             & \ce{P4}                 & Tetraphosphorus                           & 4  & 600 - 360   & 6  & \cite{93Ed}  \\
        \ce{C6H8}              & (3E)-hexa-1,3,5-triene           & 14 & 3099 - 90   & 36 & \cite{94ChVe}                                                           & \ce{PCl3}               & Trichlorophosphane                        & 4  & 504 - 198   & 6  & \cite{65MiScFa} \\
        \ce{CCl2F2}            & Dichloro(difluoro)methane        & 5  & 1161 - 261  & 9  & \cite{14EvSiMe, 90GiGaBa, 79GiGaFr, 14RoMeMc, 82JoMoCh}                 & \ce{PF3}                & Trifluorophosphane                        & 4  & 891 - 347   & 6  & \cite{14Na} \\
        \ce{CCl4}              & Tetrachloromethane               & 5  & 793 - 218   & 9  & \cite{02ChVe, 06ChRa, 14GaWeVa}                                         & \ce{$^{\textit{3}}$PH}  & $\lambda^{1}$-phosphane                   & 2  & 2276 - 2276 & 1  & \cite{96RaBe} \\
        \ce{CF4}               & Tetrafluoromethane               & 5  & 1283 - 435  & 9  & \cite{17CaGrRi}                                                         & \ce{PH3}                & Phosphine                                 & 4  & 2326 - 992  & 6  & \cite{92TaLaLe, 00FuLo} \\
        \ce{$^{\textit{2}}$CH} & Methylidyne radical              & 2  & 2732 - 2732 & 1  & \cite{95Za}                                                             & \ce{PN}                 & Azanylidynephosphane                      & 2  & 1323 - 1323 & 1  & \cite{95AhHa} \\
        \ce{CH2Cl2}            & Dichloromethane                  & 5  & 3055 - 281  & 9  & \cite{86DuNiTu}                                                         & \ce{S2}                 & Disulfur                                  & 2  & 725 - 725   & 1  & \cite{94LiNiTi} \\
        \ce{CH2N2}             & Diazomethane                     & 5  & 3184 - 408  & 9  & \cite{84VoWiYa, 90NeVoWi, 64MoPi}                                       & \ce{S2F2}               & Fluorosulfanyl thiohypofluorite           & 4  & 717 - 182   & 6  & \cite{70BrPe} \\
        \ce{CH2O2}             & Formic acid                      & 5  & 3568 - 626  & 9  & \cite{82BeMi, 06BaMaAl, 09PeVaZe, 17LuZhLi, 06BaAlMo}                   & \ce{S2O}                & Disulfur monoxide                         & 3  & 1166 - 380  & 3  & \cite{16ThMaEn, 86LiRuJo, 15MaEnZi} \\
        \ce{CH2S}              & Thioformaldehyde                 & 4  & 3024 - 990  & 6  & \cite{71JoOl, 93McBr, 08FlLaPe}                                         & \ce{SCl2}               & Chloro thiohypochlorite                   & 3  & 528 - 205   & 3  & \cite{00BiClDe} \\
        \ce{CH3Cl}             & Chloromethane                    & 5  & 3039 - 732  & 9  & \cite{11BrPeJa, 03NiFeCh}                                               & \ce{$^{\textit{2}}$SH}  & $\lambda^{1}$-sulfane                     & 2  & 2599 - 2599 & 1  & \cite{95RaBeEn} \\
        \ce{CH3F}              & Fluoromethane                    & 5  & 2998 - 1048 & 9  & \cite{92PaPaOg, 91PaOgCi, 91PaTePr, 82ChRoMi}                           & \ce{Si2H6}              & Disilane    & 8  & 2169 - 130  & 18 & \cite{57BeKe, 05LaDiHo, 03LaDiHo, 04LaDiHo_2, 92Mc_2, 04LaDiHo} \\
        \ce{CH3N}              & Methanimine                      & 5  & 3262 - 1058 & 9  & \cite{85HaDu2091, 81DukaLe, 82DuLe, 85HaDu, 85HaDu118}                  & \ce{SiH2}               & Silylene                                  & 3  & 1995 - 998  & 3  & \cite{99HiIs} \\
        \ce{CH3NO}             & Formamide                        & 6  & 3563 - 288  & 12 & \cite{09SzHo, 99McEvLa}                                                 & \ce{SiH2Cl2}            & Dichlorosilane                            & 5  & 2237 - 188  & 9  & \cite{68ChNi} \\
        \ce{CH4}               & Methane                          & 5  & 3018 - 1310 & 9  & \cite{18AmGaGe}                                                         & \ce{SiH3Cl}             & Chlorosilane                              & 5  & 2201 - 550  & 9  & \cite{84BaCa, 56NeKeLo}  \\
        \ce{CH4O}              & Methanol                         & 6  & 3682 - 199  & 12 & \cite{10TwClPe}                                                         & \ce{SiH3F}              & Fluorosilane                              & 5  & 2209 - 729  & 9  & \cite{82EsGaBu, 71RoCaHo} \\
        \ce{CH4S}              & Methanethiol                     & 6  & 3010 - 234  & 12 & \cite{16LeXuBi, 18LeXuTw, 17DuLiGa, 18LeXuGu, 21UnNaOz, 16DuZhZh}       & \ce{SiH4}               & Silane                                    & 5  & 2189 - 913  & 9  & \cite{76JoKrSu, 66WiWi, 78CaGrRo} \\
        \ce{CH5N}              & Methanamine                      & 7  & 3424 - 264  & 15 & \cite{20GuKrAs, 11GuKrHo, 57GrLo, 10GuLoKr, 20GuKr}                     & \ce{SiHCl3}             & Trichlorosilane                           & 5  & 2260 - 175  & 9  & \cite{70BuRu} \\
        \ce{CH5P}              & Methylphosphine                  & 7  & 3002 - 219  & 15 & \cite{01KiZe}                                                           & \ce{SiHF3}              & Trifluorosilane                           & 5  & 2315 - 305  & 9  & \cite{00BuMkDe, 00GnMaCo, 92HaAkMi, 71BuBiRu} \\
        \ce{CH6Si}             & Methylsilane                     & 8  & 2981 - 187  & 18 & \cite{09BoOzBa, 62Wi}                                                   & \ce{$^{\textit{1}}$CH2} & Methylene                                 & 3  & 2864 - 1352 & 3  & \cite{83PeNeOg, 89PeNeDa, 87PeNeBr} \\
        \ce{CHCl3}             & Chloroform                       & 5  & 3032 - 260  & 9  & \cite{12PrCeHo, 02PiAlHo, 00PiHoAn, 04AnAlHo, 95PaHoPi, 99PiHoAn}       & \ce{SiO}                 & Oxosilicon                               & 2  & 1241 - 1241 & 1  & \cite{03SaMcTh} \\
        \ce{CHF3}              & Fluoroform                       & 5  & 3018 - 507  & 9  & \cite{18AlBaBe, 84DuQu, 03CeCoDe}                                       & \ce{SO2}                 & Sulfur dioxide                           & 3  & 1362 - 517  & 3  & \cite{05MuBr, 87GuNaUl} \\
        \ce{Cl2}               & Molecular chlorine               & 2  & 559 - 559   & 1  & \cite{75DoHo}                                                           & \ce{SO3}                 & Sulfur trioxide                          & 4  & 1391 - 497  & 6  & \cite{01MaBlSa, 02BaChMa} \\
        \ce{Cl2O}              & Chloro hypochlorite              & 3  & 686 - 300   & 3  & \cite{65RoPi, 96XuMcBu}                                                 & \ce{SOCl2}               & Thionyl dichloride                       & 4  & 1251 - 194  & 6  & \cite{17RoDhCu, 15MaMoPi, 54MaLa} \\
        \ce{ClCN}              & Carbononitridic chloride         & 3  & 2249 - 381  & 4  & \cite{94SaDuBl}                                                         & \ce{trans-C2H4Cl2}       & 1,2-dichloroethane                       & 8  & 3018 - 137  & 18 & \cite{75MiShHa}  \\
        \ce{ClF}               & Chlorine fluoride                & 2  & 773 - 773   & 1  & \cite{86BuScJa}                                                         & \ce{$^{\textit{3}}$CH2}  & Methylene                                & 3  & 3213 - 963  & 3  & \cite{86MaMc, 88JeBu} \\
        \ce{ClF3}              & Trifluoro-$\lambda^{3}$-chlorane & 4  & 751 - 328   & 6  & \cite{70SeClHo}                                                         &                          &                                          &    &             &    & \\
    \bottomrule
    \end{tabular}
    }
\end{table}

VIBFREQ1295 was built upon data collected from six popular vibrational frequency databases available in the literature that have all been previously used to calculate scaling factors, indicating the suitability of the collated molecules for this purpose. \Cref{tab:parent_dbs} outlines the databases considered, describing their original source of data when compiling the experimental fundamental frequencies. We chose these databases based on the heterogeneity of the molecules considered to guarantee minimal overlap of molecules between the different sets. In \Cref{fig:datasets} we present the six databases considered, showcasing the size of each database as indicated by the notation \textit{(x)f/(y)mol}, where $x$ and $y$ represent the total number of frequencies and molecules, respectively.  It is worth mentioning some other databases of critically evaluated experimental data, especially the Computational Chemistry Comparison and Benchmark Database (CCCBDB) \cite{20Ru}, the compilation of experimental vibrational frequencies by Shimanouchi \cite{77Sh}, and the database collated by Huber and Herzberg \cite{79HuHe}. Though these bigger databases contain additional molecules and data that could be used to expand our new database, we chose to align more closely with the heritage of previous database collation for vibrational frequency scaling factors, which were often extracted from these larger databases.

\begin{table*}
    \centering
    \caption{Parent databases considered in the development of VIBFREQ1295. }
    \label{tab:parent_dbs}
    \scalebox{0.75}{
    \begin{tabular}{p{0.1\linewidth} p{0.2\linewidth}  p{0.2\linewidth}  p{0.005\linewidth} p{0.002\linewidth} p{0.45\linewidth}}
    \toprule
        \mc{1}{c}{Database}  & \mc{1}{c}{Alias} & \mc{1}{c}{Reference}  & \mc{1}{c}{Freqs} & \mc{1}{c}{Mols} & \mc{1}{c}{Data Origin}\\
        \midrule
            04WiMo   & 1306f/65mol  &  \citet{04WiMo}   & 1306 & 65            & Largely based on the Computational Chemistry Comparison and Benchmark   Database (CCCBDB) \cite{20Ru}.                                                                                 \\
            93PoScWo & 1064f/122mol &  \citet{93PoScWo}                   & 1064 & 122           & Frequencies collected from the herculean work of Shimanouchi \cite{77Sh}. Widley   implemented in the calculation of scaling factors as evidenced by  Scott and Radom \cite{96ScRa}.                            \\
            21UnNaOz & 842f/157mol  &  \ \citet{21UnNaOz}                 & 824  & 157           & Largely based on the Computational Chemistry Comparison and Benchmark   Database (CCCBDB) \cite{20Ru}.                                                                                 \\
            93HeHo   & 510f/42mol   &  \citet{93HeHo}     & 510  & 42            & Largely based on the Computational Chemistry Comparison and Benchmark   Database (CCCBDB) \cite{20Ru}.                                                                                 \\
            15KeBrMa & 119f/30mol   &  \citet{15KeBrMa}                   & 119  & 30            & Combination of experimental and computational data taken from the CCCBD   set \cite{20Ru} and the Huber and Herzberg compilation \cite{79HuHe}, as   well as from individual studies. \\
            10AlZhZh & 50f/15mol    &  \citet{10AlZhZh}                   & 50   & 15            & Based on the NIST Web book \cite{NIST} and the CCCBD set \cite{20Ru}. \\
    \bottomrule
    \end{tabular}}
\end{table*}

As displayed in \Cref{fig:datasets}, not all molecules and frequencies from the former databases were included into VIBFREQ1295. Instead, we curated the initial data by considering only molecules fulfilling three main criteria:

\begin{compactenum}
    \item Rigid molecules with only one energetically-relevant low-lying conformer (this was assessed using the CREST \cite{21GrBoHa} and CENSO \cite{21GrBoHa} packages for conformer sampling with default values);
    \item Small- to medium-sized molecules with no more than 14 total atoms and 6 non-hydrogen atoms; and
    \item Molecules with easily accessible and reliable gas-phase experimental fundamental frequencies reported in the literature. 
\end{compactenum}

The light-blue areas connecting the nodes in \Cref{fig:datasets} schematically represent the number of molecules and frequencies from each database that have been considered into VIBFREQ1295, along with the percentages they represent in their parent database. In most cases, we included $>$\,60\,\% of the molecules and frequencies from the parent databases, except for the 1306f/65mol (04WiMo) \cite{04WiMo} set, where only 37.7\,\% of the data was considered. The light-red areas in the figure, on the other hand, represent the proportion of data that was excluded from our database. This data mostly consists of a combination of molecules that (1) exceed the 14 total atoms criteria, (2) have incomplete frequency assignments, and (3) rely on experimental fundamental frequencies recorded in liquid-phase or with matrix isolation techniques. Note that despite disregarding this data from the database, VIBFREQ1295 is still large in terms of the number of experimental fundamental frequencies considered. 

To avoid carrying forward potential misassignments in the fundamental frequencies, we performed a thorough literature search to search for potential updates to the original experimental fundamental frequencies for all molecules in our database. 
Frequencies from newer publications replaced the original data when either (1) the new publication had data at a significantly higher resolution (usually rovibrational resolved vs unresolved) and (2) the new publication explicitly discussed and justified the reassignments of the experimental data (often with computational or extra experimental evidence).

\begin{figure}
    \centering
    \includegraphics[width=0.7\textwidth]{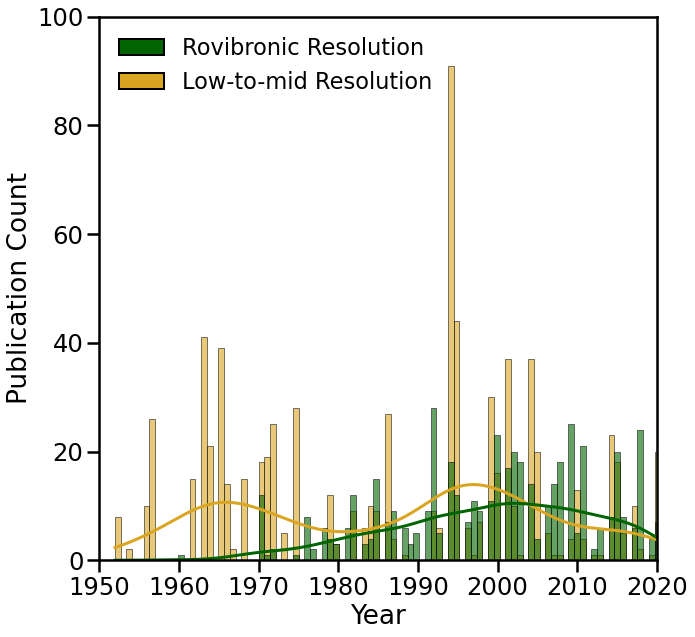}
    \caption{Distribution of the year of publication for the experimental data compiled in the present work. Green and yellow are used to distinguish between publications with rovibronic and low-to-mid resolution, respectively.}
    \label{fig:publication_year}
\end{figure}

\Cref{fig:publication_year} presents a histogram of the compiled experimental data as a function of the year of publication. We observe that, whilst the time-length of publications spans between the 1950s to 2020, 39.3\,\% of the collated data (i.e., 509 fundamental frequencies) have been published since 2000.

\begin{figure}
    \centering
    \includegraphics[width=0.7\textwidth]{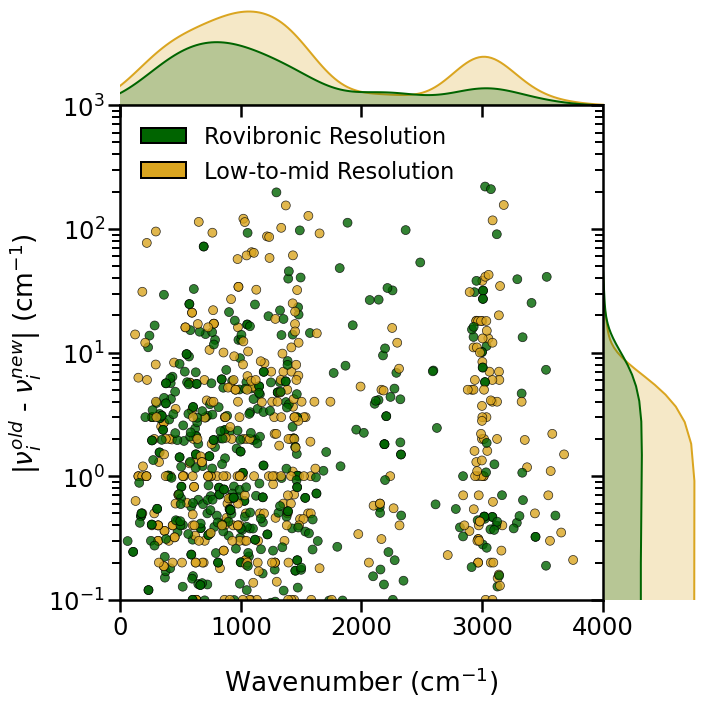}
    \caption{Absolute deviation (in logarithmic scale) between the former ($\nu_{i}^{old}$) and updated ($\nu_{i}^{new}$) experimental fundamental frequencies ($|\nu_{i}^{old} - \nu_{i}^{new}|$) in VIBFREQ1295. The plots on the top and right-side of the figure illustrate the density distribution of the data across the whole frequency and absolute deviation ranges, respectively. Differences below 10$^{-1}$\,\cm{} are not displayed in the figure.}
    \label{fig:freq_diff}
\end{figure}

\begin{table}
    \centering
    \caption{Frequency difference, $|\nu_{i}^{old} - \nu_{i}^{new}|$ (in \cm{}), breakdown between the former $\nu_{i}^{old}$ and updated $\nu_{i}^{new}$ values in VIBFREQ1295.}
    \label{tab:freqs_database}
    \scalebox{1}{
    \begin{tabular}{lcc}
        \toprule
        \mc{1}{c}{$|\nu_{i}^{old} - \nu_{i}^{new}|$} & \mc{1}{c}{Percentage (\%)} & \mc{1}{c}{No. of Frequencies} \\
        \midrule
        Identical  & 30.4 & 394 \\
        (0 -- 1]   & 34.4 & 445 \\
        (1 -- 10]  & 24.7 & 320 \\
        (10 -- 50] & 8.11 & 105 \\
        $>$ 50     & 2.39 & 31 \\ 
        \bottomrule
    \end{tabular}}
\end{table}

\begin{table*}
    \centering
    \caption{Molecules with an absolute frequency difference, i.e., $|\nu_{i}^{old} - \nu_{i}^{new}|$, larger than 50\,\cm{}. The potential reason behind these discrepancies is provided in the table, as well as the references to the former and updated frequency assignments.}
    \label{tab:large_diff}
    \scalebox{0.52}{
    \begin{tabular}{lcclclcl}
        \toprule
        \mc{1}{c}{Molecule} & \mc{1}{c}{Mode} & \mc{1}{c}{Former Freq. (\cm{})} & \mc{1}{c}{Former Reference} &\mc{1}{c}{New Freq. (\cm{})} & \mc{1}{c}{New Reference} & \mc{1}{c}{$|\nu_{i}^{old} - \nu_{i}^{new}|$ (\cm{})} & \mc{1}{c}{Reason} \\
        \midrule
            \ce{BH}                           & 1                  & 2367.0                 & Fernando and Bernath \cite{91FeBe}                    & 2269.2           & Fernando and Bernath \cite{91FeBe}                     & 97.773            & Former frequency was harmonic \\ [1.5mm]
            \ce{$^{\textit{2}}$BO}                           & 1                  & 1885.0                 &  \citet{85MeDuBr}                  & 1772.9            &  \citet{85MeDuBr}                  & 112.14             & Former frequency was harmonic                  \\ [1.5mm]
            \multirow{2}{*}{\ce{C2H5F}}       & 8                  & 1048.0                 & \multirow{2}{*}{ \citet{93PoScWo}$^{*}$} & 1109.0                 & \multirow{2}{*}{Dinu \etal{} \cite{20DiZiPo}} & 61.000                  & \multirow{2}{*}{Potential missassignment}       \\
                                              & 16                 & 1108.0                 &                                   & 1172.0                 &                                   & 64.000                  &                                                \\ [1.5mm]
            \ce{C2H6S}                        & 14                 & 973.00                  & Witek and Morokuma\cite{04WiMo}$^{*}$                    & 1030.0                 & Ellwood \etal{} \cite{94ElStGe}                  & 57.000                  & Potential missassignment                        \\ [1.5mm]
            \multirow{2}{*}{\ce{C4N2}}        & \multirow{2}{*}{3} & \multirow{2}{*}{692.00} & \multirow{2}{*}{Chase \etal{} \cite{75ChCuPr}} & \multirow{2}{*}{620.00} & \multirow{2}{*}{Winther and Sch\"onhoff\cite{97WiSc}}   & \multirow{2}{*}{72.000} & \multirow{2}{*}{Potential missassignment}       \\
                                              &                    &                      &                                   &                      &                                   &                     &                                                \\ [1.5mm]
            \multirow{2}{*}{\ce{CH2S}}        & 5                  & 3180.7              & \multirow{2}{*}{ \citet{11YaYuRi}} & 3024.6            & \multirow{2}{*}{Johns and Olson\cite{71JoOl}}   & 156.07            & \multirow{2}{*}{Former frequency was harmonic} \\ 
                                              & 1                  & 3088.2              &                                   & 2971.0            &                                   & 117.21            &                                                \\ [1.5mm]
            \ce{CH4O}                         & 12                 & 250.00                  & Healy and Holder \cite{93HeHo}$^{*}$                    & 199.80                & Twagirayezu \etal{} \cite{10TwClPe}                  & 50.200                & Potential missassignment                        \\ [1.5mm]
            \ce{CH5P}                         & 5                  & 1238.0                 & \ \citet{21UnNaOz}$^{*}$                  & 1296.2               & Kim and Zeroka \cite{01KiZe}                    & 58.200                & Potential missassignment                        \\ [1.5mm]
            \ce{CSCl2}                        & 3                  & 220.00                 & \citet{93PoScWo}$^{*}$                  & 297.00                  & Frenzel \etal{} \cite{70FrBlBe}                  & 77.000                  & Potential missassignment                        \\ [1.5mm]
            \ce{$^{\textit{2}}$HCO}                          & 1                  & 2488.0                 & \citet{93PoScWo}$^{*}$                  & 2434.5            & Dane \etal{} \cite{88DaLaCu}                  & 53.522             & Potential missassignment                        \\ [1.5mm]
            \multirow{2}{*}{\ce{HO2}}         & 2                  & 1489.0                 & \multirow{2}{*}{\ \citet{21UnNaOz}$^{*}$} & 1391.8           & \citet{92BuHaHo}                  & 97.246            & \multirow{2}{*}{Potential missassignment}       \\
                                              & 3                  & 1295.0                 &                                   & 1097.6            & Nelson and Zahniser \cite{91NeZa}                    & 197.38            &                                                \\ [1.5mm]
            \multirow{10}{*}{\ce{c-C3H3NO}\_1} & 13                 & 764.00                  & \multirow{10}{*}{Witek and Morokuma\cite{04WiMo}$^{*}$}  & 857.10                & \multirow{10}{*}{Robertson \cite{05Ro}}    & 93.100                & \multirow{10}{*}{Potential missassignment}      \\
                                              & 12                 & 1021.0                 &                                   & 900.20                &                                   & 120.80               &                                                \\
                                              & 16                 & 651.00                  &                                   & 764.90                &                                   & 113.90               &                                                \\
                                              & 4                  & 1653.0                 &                                   & 1561.1               &                                   & 91.900                &                                                \\
                                              & 5                  & 1560.0                 &                                   & 1432.6               &                                   & 127.40               &                                                \\
                                              & 6                  & 1432.0                 &                                   & 1370.9               &                                   & 61.100                &                                                \\
                                              & 7                  & 1373.0                 &                                   & 1218.3               &                                   & 154.70               &                                                \\
                                              & 8                  & 1217.0                 &                                   & 1130.0                 &                                   & 87.000                  &                                                \\
                                              & 10                 & 1089.0                 &                                   & 1024.20               &                                   & 64.800                &                                                \\
                                              & 11                 & 1033.0                 &                                   & 919.50                &                                   & 113.50               &                                                \\ [1.5mm]
            \multirow{2}{*}{\ce{c-C4H4N2}}    & 3                  & 1335.0                 & \multirow{2}{*}{Witek and Morokuma\cite{04WiMo}$^{*}$}   & 1233.0                 & \multirow{2}{*}{Hewett \etal{} \cite{94HeShBr}} & 102.00                 & \multirow{2}{*}{Potential missassignment}       \\
                                              & 17                 & 1235.0                 &                                   & 1149                 &                                   & 86.000                  &                                                \\ [1.5mm]
            \multirow{2}{*}{\ce{$^{\textit{1}}$CH2}}     & 1                  & 3026.0                 & \multirow{2}{*}{ \citet{93PoScWo}$^{*}$} & 2806.1              & Petek \etal{} \cite{89PeNeDa}                  & 219.93              & \multirow{2}{*}{Potential missassignment}       \\
                                              & 3                  & 3074.0                 &                                   & 2864.5               & Petek \etal{} \cite{83PeNeOg}                  & 209.50               &                                                \\ [1.5mm]
            \multirow{2}{*}{\ce{$^{\textit{3}}$CH2}}     & 2                  & 1056.0                 & \multirow{2}{*}{ \citet{93PoScWo}$^{*}$} & 963.10             & Marshall and McKellar \cite{86MaMc}                    & 92.901             & \multirow{2}{*}{Potential missassignment}       \\
                                              & 3                  & 3123.0                 &                                   & 3213.5               & Jensen and Bunker \cite{88JeBu}                    & 90.500                &     \\                                          
        \bottomrule
    \end{tabular}}
        \begin{tablenotes}
            \item[] \scriptsize{(*) Note that the compilation paper, i.e., the parent vibrational frequency database, is referenced when the original experimental publication was not available.}
        \end{tablenotes}
\end{table*}

\Cref{fig:publication_year} also highlights the experimental resolution of the data, a good indication of likely uncertainty in the reported experimental figure. 497 of the collected frequencies (38\,\%) have rovibronic resolution, whereas the rest of the database consists of fundamental frequencies recorded at low- to mid-resolution.

To understand the importance of the data update, \Cref{fig:freq_diff} presents the absolute difference (in logarithmic scale) between the frequencies reported in the parent databases ($\nu_{i}^{old}$), and the frequencies collected from our new literature search ($\nu_{i}^{new}$), i.e., $|\nu_{i}^{old} - \nu_{i}^{new}|$. The plots on the top and right-side of the figure illustrate the density distribution of the data in each axis. Large errors concentrate in the region below 1500\,\cm{} (the low-frequency region) and around 3000\,\cm{} where the C--H stretching frequencies dominate; however, these are also the regions were more data is available.

Interestingly, the resolution at which the fundamental frequencies have been recorded does not significantly influence the similarity between the former and updated frequencies.

\Cref{tab:freqs_database} presents the frequency difference breakdown between the former and updated values. Note that the first row in the table corresponds to frequencies where no updated experimental values were found in the literature (this data is not displayed in \Cref{fig:freq_diff}). The last row in the table, on the other hand, shows the number of frequencies for which the difference between the former and updated values differs by more than 50\,\cm{}. We can attribute two main factors to the large discrepancies in these cases: (1) former data corresponding to experimentally-derived harmonic frequencies rather than fundamental frequencies (VIBFREQ1295 consists of experimental fundamental frequencies only), and (2) potentially unnoticed missasignments in the parent databases. In \Cref{tab:large_diff} we present the former and updated frequencies for these molecules, along with the absolute frequency difference $|\nu_{i}^{old} - \nu_{i}^{new}|$, and the references to the original and updated experimental publications.

\section{Database Overview}
\label{sec:database_overview}

\subsection{Chemical Space and Molecular Classes}

Understanding the chemical nature underpinning the database is essential for delimiting the scope of its applications. In this section, we will compare the chemical composition of the molecules in VIBFREQ1295 with those from four benchmark databases in computational quantum chemistry: the popular general-chemistry sets MGCDB84 \cite{17MaHe} and GMTKN55 \cite{17GoHaBa}, both containing reference values for thermochemistry, isomerisation energies and kinetics; as well as the specialised Dip152 \cite{18HaHe1} and Pol132 \cite{18HaHe2} databases, storing benchmark dipole moments and static polarisabilities for small molecules, respectively. For the sake of completeness, we also make comparisons against the popular 1064f/122mol vibrational frequencies database collated by Pople \etal{} \cite{93PoScWo}

\begin{figure*}
    \centering
    \includegraphics[width=0.8\textwidth]{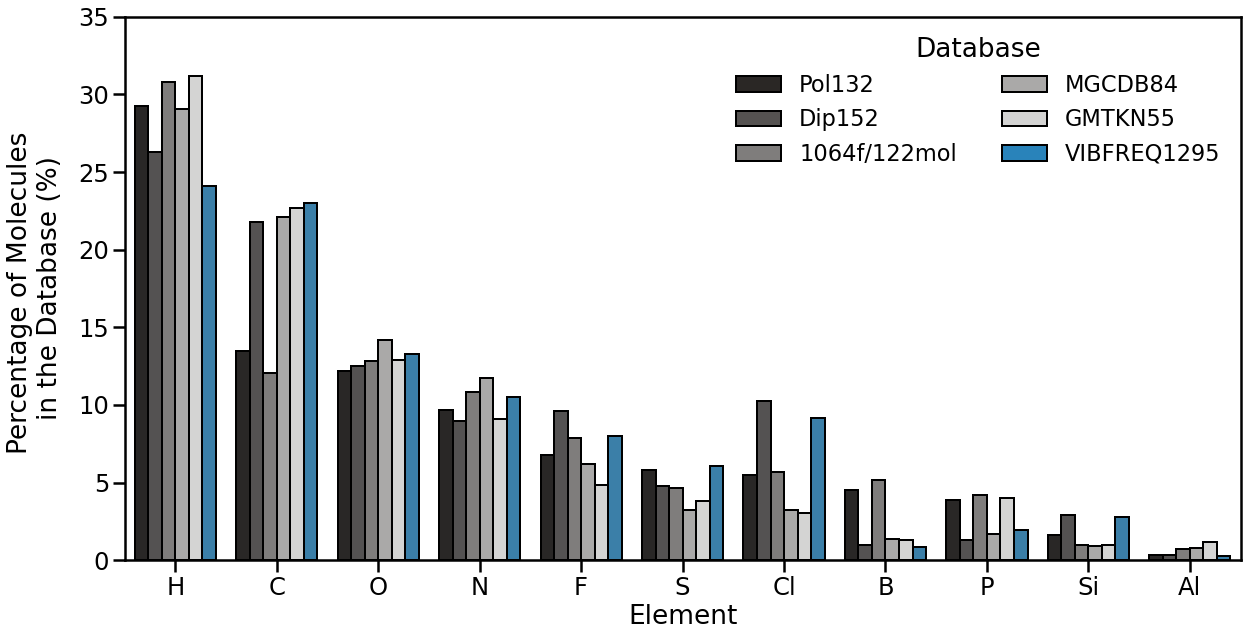}
    \caption{Comparison of the distribution of elements (in percentage of molecules) across the six benchmark databases considered: in blue, VIBFREQ1295, introduced in the present paper; and in shades of grey Pol132 \cite{18HaHe2}, Dip152 \cite{18HaHe1}, 1064f/122mol \cite{93PoScWo}, MGCDB84 \cite{17MaHe}, and GMTKN55 \cite{17GoHaBa}, respectively. Note that, apart from VIBFREQ1295, the other databases may also have molecules containing elements different than those displayed in the figure.}
    \label{fig:elements}
\end{figure*}

\Cref{fig:elements} displays the range of elements available in VIBFREQ1295, along with the percentage of molecules containing each element individually. We can observe that the element composition in the database is constrained to molecules containing nonmetal and halogen atoms, with a small percentage of molecules containing the less common B, Si and Al elements. Thus, future applications of VIBFREQ1295 should focus towards general-chemistry properties of rigid organic-like molecules only.

\Cref{fig:elements} also compares the molecular element compositions of VIBFREQ1295 with those from the other general-chemistry and specialised databases. Note that the percentages in the figure are indicative only, as the naming convention for some molecules in the databases was not consistent. All six databases share a similar trend in element composition focusing particularly in organic-like molecules containing H, C, O, N and halogens (except for the Pol132 and 1064f/122mol sets where C-containing molecules are less prominent). Even though 70.7\,\% of the data from 1064f/122mol was included into VIBFREQ1295 (see \Cref{fig:datasets}), the figure shows some significant differences in the percentages for both sets, especially for Cl-containing molecules where VIBFREQ1295 is particularly large. We can attribute this difference to the widespread of Cl-containing molecules in the other vibrational frequency databases considered.

To allow further comparisons, we categorised the list of all molecules in the six databases into three main classes:

\begin{compactenum}
    \item \textbf{CHNOPS Molecules:} molecules containing C, H, N, O, P and S only (e.g., c-\ce{C4H4O}, \ce{C2N2});
    \item \textbf{Halogens:} any molecule containing a halogen element (e.g., \ce{AlCl3}, \ce{SiH3F}); and
    \item \textbf{Others:} molecules containing any other element from the periodic table, except for halogens (e.g., \ce{BH3CO}, \ce{CH6Si}); and
\end{compactenum}

\begin{figure*}
    \centering
    \includegraphics[width=0.9\textwidth]{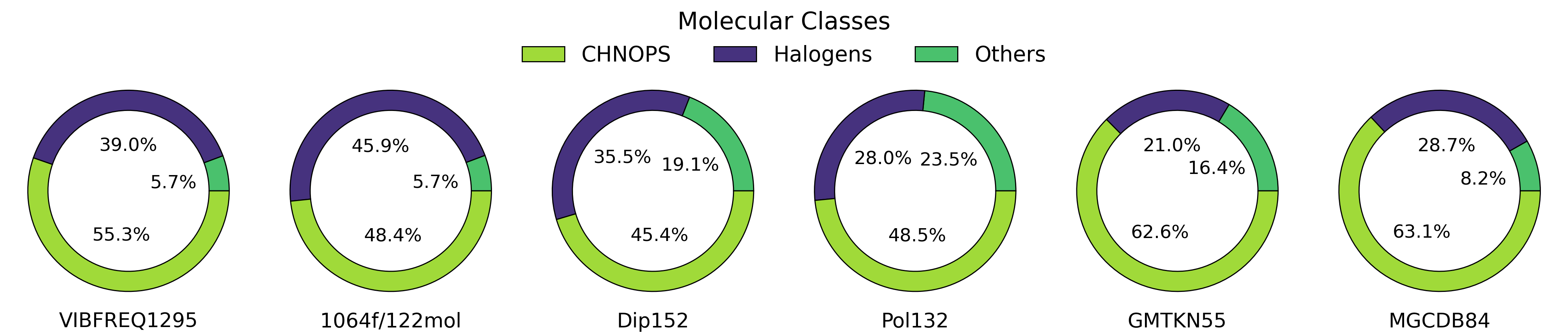}
    \caption{Proportion of molecules in the databases considered that fall within the three molecular classes defined in present work: CHNOPS, Halogens, and Others.}
    \label{fig:mol_class}
\end{figure*}

The pie charts in \Cref{fig:mol_class} show the percentages for each molecular class in the different databases considered. \textit{CHNOPS-like} molecules dominate across all databases, especially in the VIBFREQ1295, GMTKN55 and MGCDB84 sets with more than 50\,\% of the molecules falling within this class. Molecules belonging to the \textit{Others} class are scarce in VIBFREQ1295 (only 5.7\,\% of the database), mostly due to the absence of species containing B, Si and Al (see \Cref{fig:elements}), similarly to the 1064f/122mol and MGCDB84 sets which also seem to underestimate this molecular class. Nonetheless, VIBFREQ1295 is notably rich in halogenated compounds (almost 40\,\% of the database), which correlates with the large percentage of Cl-containing molecules in the database. As both \textit{CHNOPS-like} and halogenated compounds share similar percentages in VIBFREQ1295, we do not expect significant biases towards a particular class in future applications (see Section \ref{sec:qc-calcs}). However, we caution the user about the potential overestimation of halogenated compounds in VIBFREQ1295.

\subsection{Functional Groups}

We wrote a python code based on the RDKtid library \cite{10La} to obtain the counts of the different functional groups in our database. A comprehensive description of all functional groups can be found in Table S1 in the supplemental material for this paper. 

\Cref{fig:fgroups} illustrates some of the functional groups present in VIBFREQ1295, along with their indicative counts in parenthesis. In line with the predominance of F- and Cl-containing species, halogen-containing functional groups dominate in the database,
particularly due to the ubiquitous presence of the E--X group (where $E$ represents any element apart from C, and $X$ is a halogen), having the largest count across all functional groups. Hydrocarbon groups are also widespread in the database, with a large proportion of molecules containing alkane- and alkene-like groups. Note that the variety of O-containing groups is extensive, with multiple functional groups containing pairs of heteroatoms including O, e.g., R--NO$_{3}$ (nitrate) or R$_{1}$--SO--R$_{2}$ (sulfoxide).

\begin{figure*}
    \centering
    \includegraphics[width=0.9\textwidth]{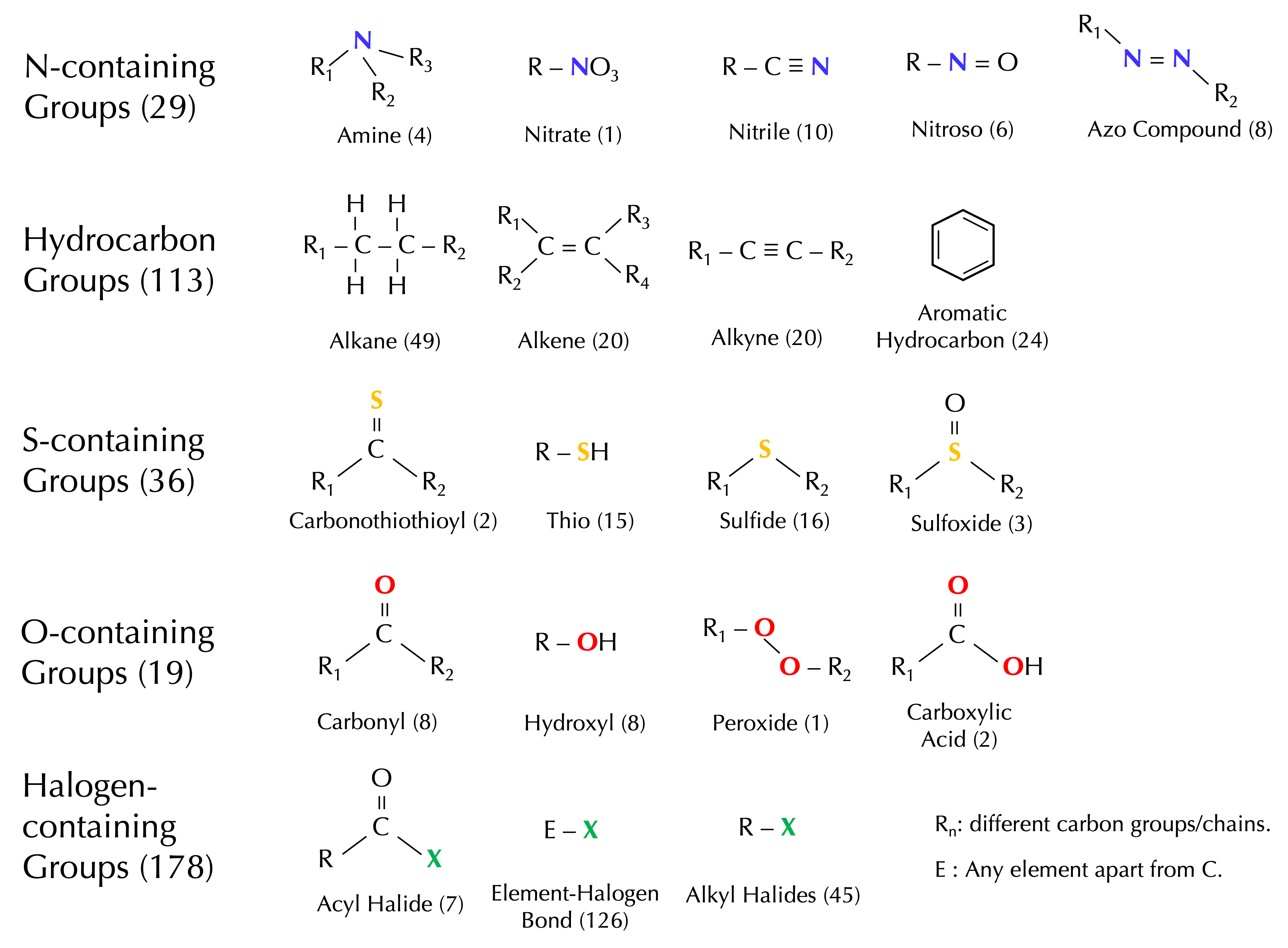}
    \caption{Schematic representation of popular functional groups in organic chemistry present in VIBFREQ1295. The functional groups are sorted based on their element composition. The numbers in parenthesis are indicatives of the occurrence of the functional groups in the database. $R_{n}$ is used to denote C-containing groups or chains, and $E$ represents any element from the periodic table apart from C.}
    \label{fig:fgroups}
\end{figure*}

\subsection{Vibrational Modes}

Functional groups are not only responsible for driving chemical reactivity, but also help distinguishing between molecules based on their vibrational motions. As functional groups correspond to well-defined regions of the molecule containing specific atoms (e.g., C=O group), it is straightforward to identify the localised vibrations of those functional groups in the infrared spectrum (e.g., C=O stretch at $\sim$1700\,\cm{}). 

In VIBFREQ1295, we provide approximate descriptions for all vibrational modes in the database, based on the vibrational motions from our quantum-chemistry calculations (see Section \ref{sec:qc-calcs}). Note that these descriptions are indicative only; Highly-coupled modes and descriptions in the low-frequency range (usually below 1,000\,\cm{}) are less accurate due to the large number of atoms and bonds involved in the vibration.

\begin{figure}
    \centering
    \includegraphics[width=0.7\textwidth]{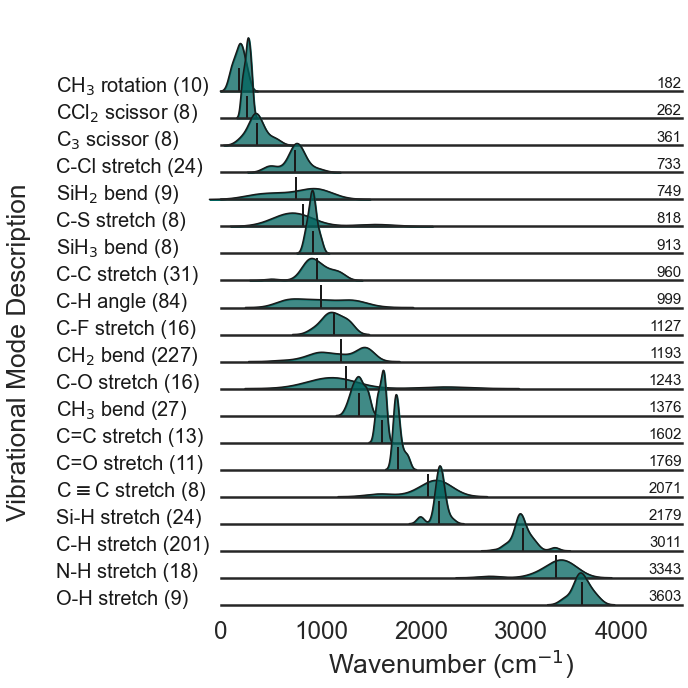}
    \caption{Density distribution of the 20 most popular vibrational modes in VIBFREQ1295. The numbers in parenthesis correspond to the number of entries of each vibrational mode in the database. The central frequencies for the functional group are displayed in the right-most side in the figure, as well as visually represented as vertical lines in the corresponding density distribution.}
    \label{fig:vmodes}
\end{figure}

\Cref{fig:vmodes} presents the density distribution for the 20 most common vibrational modes in the database. The numbers in parenthesis indicate the counts for each vibrational mode, with central frequencies displayed in the right-most side of the figure. We can observe that the database has a good coverage in terms of vibrational modes, spanning between the far- (10 -- 400\,\cm{}) and mid-infrared (400 -- 4,000\,\cm{}) regions. Popular modes like the C--H and O--H stretches are well-defined in the database, with their expected central frequencies lying at approximately 3,000 and 3,600\,\cm{}, respectively. However, non-stretch modes were more difficult to define uniquely and we kept the terms fairly generic - e.g., C-H angle and \ce{CH2} bend are used as classifiers for a variety of different vibrational motions (including some wagging, rocking, twisting and scissoring modes) and thus they span a wider frequency range. The figure also shows that, as expected, vibrational modes involving more than two atoms, or relatively heavy elements, e.g., S, Cl, lie in the low-frequency range of the spectrum (usually below 1,000\,\cm{}).

\subsection{Applications for this database}

It is clear that the primary use of VIBFREQ1295 is to allow benchmarking of the performance of model chemistries for harmonic frequency calculations. We are currently undertaking a benchmarking study quantifying the quality of scaled harmonic vibrational frequencies produced from a wide cross-section of density functional approximations, and double and triple-zeta basis sets. Future studies considering the performance of various anharmonic treatments, e.g., VPT2 and VSCF, are natural and important follow-ups. 

It is possible that future density functional theories or basis sets could utilise this vibrational frequency data as part of their training data, perhaps producing approximations specifically tuned to this property. However, this is less likely than for other experimental property databases because (1) frequencies are computationally expensive calculations, and (2) vibrational frequencies are not a particularly challenging property for model chemistries to predict as anharmonicity error dominates over model chemistry error very quickly \cite{21ZaMc}.

Instead, more promising applications could be directed towards machine learning algorithms. Models trained on the VIBFREQ1295 experimental data could predict fundamental frequencies of small molecules either entirely without computational quantum chemistry, or as corrections to these calculations. For instance, The RASCALL computational chemistry approach \cite{19SoPeSe} uses experimental fundamental frequencies (mostly from common organic functional groups) to predict approximate spectral data for thousands of molecules simultaneously. RASCALL could benefit from the distribution of vibrational frequencies collated here for multiple functional groups, as this data can help considering the effects of the chemical environment in the vibrational frequency for a given functional group, i.e., how the frequency for the functional group vibration shifts according to the atoms and chemical groups surrounding it (see \Cref{fig:vmodes}).

Another potential application of VIBFREQ1295 is providing a centralised repository of high-level \abinitio{} harmonic frequencies for supporting approaches seeking more accurate vibrational spectra predictions. This is the case of, for example, the hybrid approach implemented by Biczysko \etal{} \cite{10BiPaSc} where CCSD(T)-like harmonic frequencies for a given molecule are corrected by means of anharmonic calculations (usually VPT2) performed at a computationally less-demanding model chemistry choice, e.g., hybrid functionals and double-zeta basis sets. This hybrid approach has proved to deliver satisfactory performance in small-to-medium-sized molecules \cite{14BaBiBl,15BaBiPu,18BiBlPu}.

It is not only the experimental and theoretical harmonic frequencies in VIBFREQ1295 that can be practical; the set of compiled literature references for each molecule are a useful starting point for more detailed spectral data compilation such as model Hamiltonians fits or assigned rovibrational transitions. These higher-resolution data are crucial for enabling remote molecular identification and environmental characterisation in complex gaseous environments including planetary atmospheres \cite{07FuCsTe,12FuCs,21Mc}. 

\section{High-quality \textit{ab initio} Harmonic Frequencies: Comparison with Experiment}
\label{sec:qc-calcs}

\subsection{Computational Details}

To complement the experimental fundamental frequencies collated here, we also calculated high-level \abinitio{} harmonic vibrational frequencies for all 141 molecules. Initial molecular geometries were obtained from their SMILES identifiers through an automated approach using the RDKit \cite{10La} and ChemCoord \cite{17We} libraries in python.

We performed all harmonic frequency calculations using the CCSD(T)(F12*) method \cite{10HaTeKo} together with the cc-pVDZ-F12 basis set \cite{08PeAdWe,10HiPe}. We chose this model chemistry based on recommendations by Martin and Kesharwani \cite{14MaKe} and Schmitz and Christiansen \cite{17ScCh}, who found a fast basis set convergence at a reasonable computational cost for this model chemistry. Indeed, they reported errors between 3--4\,\cm{} in the computed harmonic frequencies when compared against CCSD(T) limit results.

The initial geometries for the calculations were optimised using a tight convergence criteria, with maximum force and maximum displacements smaller than 1.5x10$^{-5}$ Hartree/Bohr and 6.0x10$^{-5}$ \AA, respectively. Default values were used for the CABS singles (1.0x10$^{-8}$) and CCSD convergence (1.0x10$^{-6}$) thresholds. A template of the input file used in the calculations is provided in the supplemental material for this paper. All calculations were carried out using the MOLPRO 2020 quantum chemistry package \cite{20WeKnMa}. 

\subsection{Calculating Scaling Factors: The Test and Training Set Separation}

Scaling factors are essential for comparing calculated harmonic frequencies with experimental fundamental frequencies \cite{21ZaMc}. The standard approach usually involves fitting and testing the scaling factor using the same set of fundamental frequencies without distinguishing between independent training and test sets. 

Here, however, in line with good-practice procedures in data science, we computed our scaling factor by splitting the database into two different sets: a training set, used in the fitting of the scaling factor, and a test set to evaluate the performance of the computed scaling factor. Both the training and test sets have been assigned with similar percentage compositions in terms of the molecular classes (halogens $\sim$ 29\,\% and non-halogens $\sim$ 71\,\%), and vibrational modes (stretch $\sim$ 44\,\% and non-stretch $\sim$ 56\,\%) considered. Ensuring high similarity between the training and test sets is pivotal for extrapolating conclusions.

To analyse how sensitive the scaling factor and scaling factor's performance are to the partition of the database, we split VIBFREQ1295 into four different groups of training and test sets: 100 -- 0, 90 -- 10, 80 -- 20, and 70 -- 30, where the numbers in each pair correspond to the percentages of data from VIBFREQ1295 included in the training and test sets, respectively. The first partition group (100 -- 0) corresponds to the standard approach of using the whole set of data to fit and test the scaling factor. 

In all cases, we used the training set to calculate the corresponding scaling factor $\lambda$ as $\lambda = (\sum_{i}^{N} \omega_{i}\nu_{i})/(\sum_{i}^{N}\omega_{i}^{2})$, where $\omega_{i}$ and $\nu_{i}$ represent the calculated harmonic and experimental fundamental frequencies, respectively, and both summations run over the total number of frequencies $N$ considered in the training set. The calculated scaling factors were then used to scale the raw harmonic frequencies in the corresponding test sets, with performance judged by the root-mean-squared-error (RMSE), the standard statistical metric for assessing performance \cite{21ZaMc}, between the scaled harmonic and experimental fundamental frequencies as $\textrm{RMSE} = \sqrt{\left( \sum_{i}^{N} (\lambda \omega_{i} - \nu_{i})^{2}) \right)/N}$.

To ensure convergence in the reported metrics, we randomised the partition of the training and test sets over 100 cycles, calculating each time the scaling factor and corresponding RMSE. We present the average results with their standard deviations in \Cref{tab:performance} for both the training and test sets, separately, across the different partition groups.

\begin{table}
    \centering
    \caption{Training and test sets groups used in the calculation of scaling factors for the CCSD(T)(F12*)/cc-pVDZ-F12 harmonic frequencies. Numbers in parenthesis are one standard deviation in the last digit of the reported value, calculated from 100 different  training/test partitions of full dataset.}
    \label{tab:performance}
    \scalebox{1}{
    \begin{tabular}{ccccc}
        \toprule
        \multirow{2}{*}{Scaling Factor} & \multirow{2}{*}{\%} & \mc{1}{c}{Training Set} & \multirow{2}{*}{\%}  & \mc{1}{c}{Test Set} \\
        \cmidrule(r){3-3} \cmidrule(r){5-5}
         & & \mc{1}{c}{RMSE (\cm{})} & & \mc{1}{c}{RMSE (\cm{})}  \\
        \midrule
        
            0.9617     & 100 & 33       & 0   & -      \\
            0.9618(2)  & 90  & 33 (1)   & 10  & 31 (6) \\
            0.9618(3)  & 80  & 33 (1)   & 20  & 32 (5) \\
            0.9617(3)  & 70  & 33 (2)   & 30  & 32 (4) \\
            
        \bottomrule
    \end{tabular}}
\end{table}

\Cref{tab:performance} shows that virtually the same metrics are obtained for both the training and test sets across the different partition groups, ensuring high similarity between both sets, and thus negligible overfitting in the computed scaling factors. This implies that the metrics obtained from our analysis represent a fair indication of our model chemistry and scaling factor performance. Looking at the different partition groups in \Cref{tab:performance}, there is no major differences in the statistical figures reported; however, the overall small standard deviation in the test set for the 70 -- 30 split highlight as a good compromise for future calculations. Therefore, further discussions in this paper will focused on calculations using the 70 -- 30 split (with metrics hereafter referred as the global scaling factor), unless otherwise noted.

It is worth noting that the global scaling factor optimised here for the CCSD(T)(F12*)/cc-pVDZ-F12 model chemistry (0.9617(3)) is essentially identical to the scaling factor of 0.9627 found by Kesharwani \etal{} \cite{15KeBrMa} for the CCSD(T)(F12*)/cc-pVQZ-F12 model chemistry. This further validates our previous finding suggesting convergence of the scaling factor at quite moderate computational levels \cite{21ZaMc}, e.g., hybrid functional with double-zeta basis sets, thus suggesting the potential implementation of universal scaling factors for model chemistries above the hybrid/double-zeta level.

\subsection{Robust Error Quantification: Comparing Theory with Experiment}

\begin{figure*}
    \centering
    \includegraphics[width=1\textwidth]{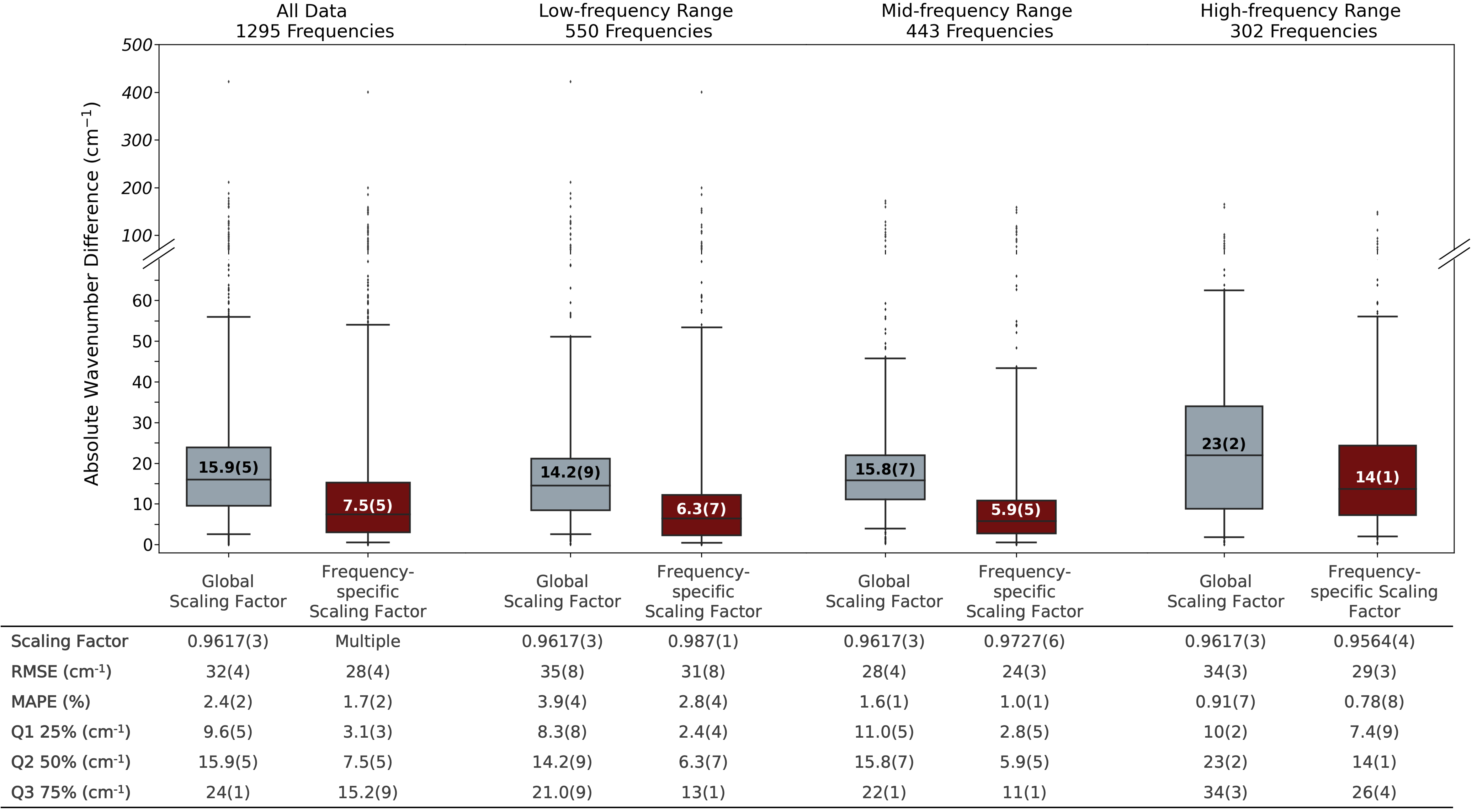}
    \caption{Absolute wavenumber difference (\cm{}) between scaled harmonic and experimental fundamental frequencies using global (grey), and frequency-range-specific (maroon) scaling factors. The numbers inside the boxes represent the median values for each classification. A scale discontinuity in the y-axis (at approximately 70\,\cm{}) has been imposed to highlight both the box and whiskers, and the outliers in the data. The table at the bottom presents the scaling factor, RMSE, Mean Absolute Percentage Error (MAPE), as well as the first, second (median), and third quartiles for the data distribution in each classification individually. The bottom and top whiskers in the figure encapsulate 5\,\% and 95\,\% of the data, respectively. Numbers in parenthesis are one standard deviation in the last digit of the reported values as calculated over 100 cycles of the 70\%/30\% training/test partitioning of the dataset.}
    \label{fig:freq_range}
\end{figure*}

The conventional approach for quantifying error in harmonic frequency calculations requires comparing experimental fundamental frequencies against calculated harmonic frequencies that have been scaled using a global scaling factor (e.g., 0.9617(3) in \Cref{tab:performance}). However, as noted by the early work of Scott and Radom \cite{96ScRa}, this procedure fails in properly describing low-frequency vibrations as it inherently gives more weight to frequencies lying at the upper end of the vibrational spectrum. Instead, alternative approaches have opted for calculating scaling factors for different frequency regions leading to differences in performance \cite{01HaVeSc, 04SiBoGu, 05AnUv, 07MeMoRa, 08AnGoJo, 11LaBoHa, 12LaCaWi, 16ChRa, 17Cha}.

The standard cutoff for distinguishing between low- and high-frequency vibrations is usually set at 1,000\,\cm{} \cite{04SiBoGu}. However, based on a detailed analysis examining the effect of multiple frequency thresholds (further explained in the supplemental material), we defined three different regions of importance: low-frequency (<\,1,000\,\cm{}), mid-frequency (1,000 -- 2,000\,\cm{}), and high-frequency ($\geq$\,2,000\,\cm{}) vibrations. 

In \Cref{fig:freq_range} we present the absolute wavenumber difference (in \cm{}) between the scaled harmonic and experimental fundamental frequencies for the defined frequency regions using the following scaling factors: in grey, using the global scaling factor (0.9617(3)) reported in \Cref{tab:performance}; and in maroon, using scaling factors optimised for the low- (0.987(1)), mid- (0.9727(6)), and high-frequency (0.9564(4)) regions. Note that the frequency-range-specific scaling factors were calculated by only considering frequencies falling within each range, e.g., the scaling factor for the low-frequency region was calculated by considering frequencies below 1,000\,\cm{} only. The number of frequencies in each frequency region is displayed at the top of the figure. The left-most boxes compare the performance of using the global and frequency-range-specific scaling factors in the whole set of data, whereas the rest of the figure compares performance for each frequency range individually. The corresponding scaling factors, RMSEs, mean absolute percentage errors (MAPE), as well as we the first, second (also displayed inside each box), and third quartiles are presented in the table at the bottom of the figure.

Overall, the figure shows a significant improvement in the absolute wavenumber difference when introducing frequency-range-specific scaling factors in the calculations (medians of 15.9(5) vs 7.5(5)\,\cm{} for the global and frequency-range-specific scaling, respectively). This result suggests that the error in the scaled harmonic frequencies is not constant, but changes gradually with the frequency region and, thus, justifies the implementation of frequency-range-specific over global scaling factors to achieve superior performance. We can confirm this variation by looking at the spread of the absolute wavenumber differences for all three designated frequency regions, where it is evident that the errors generally increase from low- to high-frequency vibrations.

The relatively high errors in the high-frequency range (when using both the global and frequency-range-specific scaling factors) can be explained by inspecting the composition of our database. A closer look into \Cref{fig:vmodes} shows that the high-frequency vibrations ($\geq$\,2,000\,\cm{}) in VIBFREQ1295 are mostly dominated by stretching-like modes, e.g., C--H, O--H, and N--H stretches, which are inherently more anharmonic than other vibrational modes in the database. This is because stretching modes can naturally lead to bond dissociation, a process not contemplated under the harmonic approximation and thus better represented by anharmonic approaches, while most non-stretching modes are harmonically bound. We can further justify this finding by noting the differences in scaling factors between the three designated frequency regions, where the scaling factor decreases from the low- (0.987(1)) and mid- (0.9727(6)), to the high-frequency (0.9564(4)) range (which also supports splitting the vibrational spectrum in three instead of two frequency regions). 

Another potential aspect to examine from this compilation of data is whether the reported metrics in \Cref{fig:freq_range} are affected by the type of molecules (halogens and non-halogens), or vibrational modes (stretches and non-stretches) in the database. We opted for excluding this analysis from the main manuscript to avoid over-complicating the narrative; however, we encourage the interested reader to look into the detailed description in the supplemental material. Instead, we summarise our findings into three main points: 

\begin{compactenum}
    \item developing scaling factors specifically tuned for halogen and non-halogen molecules is unnecessary as akin performance in the scaled harmonic frequencies is achieved when utilising the global scaling factor (0.9617(3));
    \item despite the large counts of halogen-containing molecules in VIBFREQ1295 (see \Cref{fig:elements} and \Cref{fig:mol_class}), applications of this database are not likely to be biased towards a particular molecular group as similar scaling factors and statistical metrics were found for halogen-containing and non-halogen-containing molecules; and
    \item using mode-specific scaling factors results in an overall improvement in the scaled harmonic frequencies over using a global scaling factor, potentially suggesting the implementation of different scaling factors whenever a particular vibrational mode is considered, as suggested in previous approaches \cite{83PuFoPo, 95PuRa, 98BaJaPu,07FeHoKo,07BoFeFe, 08BoDrFe,10Bo, 10BoDrFe, 21ZaMc}. Nonetheless, stretching and non-stretching modes can be generally assigned to the high and low/mid-frequency ranges of the vibrational spectrum, respectively (see \Cref{fig:vmodes}), implying that both mode-specific and frequency-range-specific scaling factors are accounting for the same corrections and could be potentially interchanged.
\end{compactenum}

In light of these findings and noting that sorting vibrational frequencies based on their frequency position is more straightforward and reliable than assigning vibrational modes from quantum-chemistry calculations (especially for highly-coupled modes), we encourage the use of frequency-range-specific (low-, mid-, and high-frequency ranges) scaling factors to achieve superior performance in harmonic frequencies calculations.

Despite the overall satisfactory performance of introducing frequency-range-specific scaling factors in harmonic frequency calculations, we also found some scaled harmonic frequencies with significant deviations from the experimental values (data points above the whiskers in  \Cref{fig:freq_range}). The reason behind these large discrepancies is unclear to us as no evident trend is observed in terms of the vibrational mode type, frequency range, or molecular geometry for these frequencies. It might be the case that these frequencies correspond to highly couple modes involving different vibrations simultaneously, which could reach the limits of accuracy within the harmonic approximation; unfortunately, however, there is not a straightforward approach to anticipate this behaviour given a random molecule.

We find open-shell molecules have generally larger errors than closed shell molecules, e.g. of the 10 open-shell species, 7 have at least one frequency with calculation error greater than 50 \cm{}. This result is in alignment with previous studies that have shown open-shell systems to be challenging for routine computational quantum chemistry calculations \cite{09BrLiGi,12TeAr,15YuTr,16YuLiTr,18Hs}. However, it is unclear whether the larger errors in these systems arise purely from increased errors in the computational chemistry method (e.g. from increased multi-reference character in these molecules) or from an increased error in the harmonic approximation for open-shell systems. 

As the frequencies with large deviations only account for approximately 5\,\% of the data distribution (top whiskers lie at 95\,\%), we expect their occurrence to be minimal in routine calculations. Table S4 in the supplemental material lists all molecules and frequencies with deviations larger than $\pm$\,50\,\cm{} from experiment, in addition to their corresponding molecular class and vibrational mode description.

\subsection{The Anharmonicity Error}

Assuming that harmonic frequency calculations at the CCSD(T)(F12*) \cite{10HaTeKo}/cc-pVDZ-F12\cite{08PeAdWe,10HiPe} level reduce model chemistry error to a minimum, we can quantify the intrinsic error in the harmonic approximation, a.k.a. the anharmonicity error, by comparing the scaled harmonic and experimental fundamental frequencies in the database.

Using the RMSE as a metric of performance, Kesharwani \etal{} \cite{15KeBrMa} first estimated this error by scaling experimentally-derived harmonic frequencies against experimental fundamental frequencies for a dataset containing 119 frequencies from 30 molecules (see \Cref{fig:datasets}). They found that an optimal scaling factor of 0.9627 results in an RMSE of 24.9\,\cm{}. We found a very similar scaling factor with the data in our database (0.9617(3) in \Cref{tab:performance}). However, our computed RMSE 
(32(4)\,\cm{}) when implementing the global scaling approach is significantly higher than the Kesharwani \etal{} \cite{15KeBrMa} value. One potential reason for this deviation lies in the heterogeneity of both data sets, but the approximate errors in our model chemistry choice, e.g., remaining basis set incompleteness error and the lack of higher order correlation effects, could also influence on the accuracy of our results \cite{07TeKlHe,14MaKe,17ScCh}.

Using the RMSE as an estimate for anharmonicity error comes with two main limitations: (1) RMSEs are highly influenced by the presence of outliers in the data, and (2) the RMSEs are a single number that doesn't provide information about the distribution of errors. These limitations are evident in the results in \Cref{fig:freq_range} where similar RMSEs are obtained for the different frequency regions when using the global or frequency-range-specific scaling factors despite the error distribution clearly improving substantially when using the frequency-range-specific scaling factors. 

Instead of the RMSE, the median error (Q2 50\,\%) for each absolute wavenumber difference distribution visually represents a more appropriate metric for anticipating the error between the predicted scaled harmonic and experimental fundamental frequency. Thus we recommend that median errors be used instead of RMSE when quantifying anharmonicity error because medians are simple to relate to the usual error of a calculation rather than emphasising outliers like RMSEs. 

The median anharmonicity error, and its distribution computed here, sets a lower bound to the reliable performance of a faster more approximate model chemistry such as a hybrid or double-hybrid functional with a double or triple-zeta basis set. Though cancellation of errors could reduce the computed anharmonicity error, there is no physical rationale for this and thus lower errors could be unreliable.

\section{Concluding Remarks}
\label{sec:conclusions}

Routine quantum-chemistry calculations, e.g., harmonic and anharmonic frequency calculations, have the potential to enable high-throughput approaches for producing approximate spectral data for thousands of molecules of astrochemistry interest \cite{21ZaSyRo}. However, it is currently unclear what model chemistry should be used and what errors in frequency predictions should be expected; this is essential information to help assess the usefulness of these high-throughput calculations to astrochemistry. 

In order to rigorously assess the errors and quantify the minimum  error obtainable by scaling harmonic computed vibrational frequencies (without unreliable cancellation of errors), here we introduce a new benchmark database (VIBFREQ1295) for harmonic frequency calculations containing 1,295 experimental fundamental frequencies, and CCSD(T)(F12*)\cite{10HaTeKo}/cc-pVDZ-F12\cite{08PeAdWe,10HiPe} \abinitio{} harmonic frequencies for 141 molecules. The database can be found as part of the supplemental material for this publication, as well as through the Harvard DataVerse \cite{Dataverse} (\url{https://doi.org/10.7910/DVN/VLVNU7}).

VIBFREQ1295's experimental component is a robust and comprehensive compilation of contemporary experimental fundamental frequencies, providing general updates in previous molecular frequency assignments. The set of compiled literature references in the database spans across nearly 80 years of research in spectroscopy, with a significant proportion of the compiled experimental data corresponding to publications between the 1990s and 2020.  More than 30\,\% of the collated frequencies have been recorded at rovibronic resolution, with the rest of the database compiling fundamental frequencies recorded at moderate resolution. This overall update in the vibrational frequencies was crucial in identifying potentially unnoticed misassignments, frequencies recorded in liquid-phase or with matrix isolation techniques, as well as the use of harmonic rather than fundamental frequencies reported in previous benchmark databases. 

The primary application of the VIBFREQ1295 experimental data is to assess the performance of computational chemistry methodologies, most importantly determining the best density functional approximations (DFAs) and basis set combinations for routine calculation of vibrational frequencies, improving on the results compiled in Zapata Trujillo and McKemmish \cite{21ZaMc} --this work is currently underway.
VIBFREQ1295 can also be used to train machine-learning models for predicting vibrational spectra of organic-like molecules.
Further, the references for the high-resolution experimental publications in the database represent a good initial source for more detailed spectral data compilations, e.g., model Hamiltonian fits.

The compiled experimental data is complemented by new  CCSD(T)(F12*) \cite{10HaTeKo}/cc-pVDZ-F12\cite{08PeAdWe,10HiPe} \abinitio{} harmonic frequencies calculations for all molecules. The comparison of the experimental and computed data allowed an in-depth analysis of the anharmonicity error, i.e., the error between the experimental fundamental frequency and scaled harmonic frequencies. The anharmonicity error was previously estimated as 24.9\,\cm{} \cite{15KeBrMa} based on RMSE differences between experimental fundamental and scaled harmonic frequencies; however, our analysis shows that median errors are more appropriate than RMSEs in predicting performance given that RMSEs are dominated by outliers rather than typical expected results. With three frequency-range-specific scaling factors (0.987(1) for frequencies < 1000 \cm{}, 0.9727(6) for frequencies between 1000 - 2000 \cm{} and 0.9564(4) for frequencies above 2000 \cm{}), our results show a median anharmonicity error of 7.5(5)\,\cm{} (RMSE 28(4)\,\cm{}); far better than the median 15.9(5)\,\cm{} (RMSE 32(4)\,\cm{}) anharmonicity error obtained with a uniform global scaling factor of 0.9617(3). We therefore recommend implementing frequency-range-specific scaling factors in harmonic frequency calculations to achieve superior performance. Future benchmarking studies should seek to determine model chemistry choices whose performance can get as close as possible to the anharmonicity error.

\begin{acknowledgement}

This research was undertaken with the assistance of resources from the National Computational Infrastructure (NCI Australia), an NCRIS enabled capability supported by the Australian Government.

The authors declare no conflicts of interest. 

\end{acknowledgement}

\begin{suppinfo}

The compiled experimental data and calculated harmonic frequencies are presented in the article and within the supplementary information. 

Specifically, the VIBFREQ1295 database is provided as a csv file with each row containing specific information for the molecules in the database (e.g., total number of atoms and non-hydrogen atoms, and molecular classification), the experimental fundamental and calculated harmonic frequencies (with vibrational modes descriptions, symmetries, resolution, and references to the original publications), as well as references to the parent databases used in the development of VIBFREQ1295.

We also provide a PDF file detailing the acronyms in the csv file, as well as additional analysis supporting our findings described in the main manuscript.

Alternatively, this data is available through the Harvard DataVerse  \cite{Dataverse} (\url{https://doi.org/10.7910/DVN/VLVNU7}). 

\end{suppinfo}


\bibliography{VibFreqDataset}

\end{document}